\documentclass[12pt]{article}
\usepackage{amsfonts}
\usepackage{amsmath}
\usepackage{amssymb}
\usepackage{mathptmx}
\usepackage{geometry}
\usepackage{graphicx}
\usepackage{hyperref}
\usepackage{cancel}
\usepackage{textcomp}
\usepackage{tikz}
\usepackage{bm}
  \usepackage{mathrsfs}
  \usepackage{filecontents}
\usepackage{times}
\usepackage{epsfig}
\usepackage{xcolor}
\usepackage{slashed}
\usepackage{booktabs}
\usepackage{latexsym}
\usepackage{verbatim}
\usepackage{extarrows}
\usepackage{multirow}
\usepackage{rotating}
\usepackage{colortbl}
\usepackage{indentfirst}
\usepackage{float}
\definecolor{mygray}{gray}{.9}
\definecolor{intnull}{RGB}{213,229,255}
\usepackage{arydshln}
\usepackage{diagbox}
\usepackage{makecell}
\usepackage{array}
\usepackage{cite}

\usepackage{caption}
\usepackage{tikz}
\usetikzlibrary{arrows,shapes,chains}
\usepackage{graphicx, subfig}
\begin{document}
\renewcommand{\thefootnote}{\fnsymbol{footnote}}
\baselineskip=16pt
\pagenumbering{arabic}
\vspace{1.0cm}
\begin{center}
{\Large\sf Can we know  about  black  hole  thermodynamics   through  shadows?}
\\[10pt]
\vspace{.5 cm}

{Xin-Chang Cai\footnote{E-mail address: caixc@mail.nankai.edu.cn} and Yan-Gang Miao\footnote{Corresponding author. E-mail address: miaoyg@nankai.edu.cn}}
\vspace{3mm}

{School of Physics, Nankai University, Tianjin 300071, China}

\vspace{4.0ex}

{\bf Abstract}
\end{center}

We investigate the relationship between shadow radius  and microstructure for a  general static spherically symmetric black hole and confirm their close connection. In this regard, we take the Reissner-Nordstr\"om (AdS) black hole as an example to do the concrete analysis. On the other hand, we  study for the Kerr (AdS) black hole the relationship between its shadow and thermodynamics  in the aspects of phase transition and microstructure. Our results for the Kerr (AdS) black hole show that  the shadow radius $r_{\rm sh}$, the deformation parameters  $\delta _{s}$ and $k_{s}$, and the circularity deviation  $\Delta C$  can reflect the black hole thermodynamics.  In addition,  we  give the constraints to  the  relaxation  time  of  the  M$87^{*}$  black  hole  by combining its shadow data  and  the Bekenstein-Hod universal bounds when the  M$87^{*}$ is regarded as  the Reissner-Nordstr\"om or Kerr black hole. We  predict that  the minimum relaxation times  of    M$87^{*}$  black  hole  and  Sgr $A^{*}$ black hole  are approximately 3 days and  2.64 minutes, respectively.  Finally,  we draw the first graph of the minimum relaxation time  $\tau _{\rm min}$  with respect to the maximum shadow radius $ r_{\rm sh}^{\rm max}$ at different mass levels.

\newpage

\section{Introduction}

Recently, the Event Horizon Telescope (EHT) collaboration has released~\cite{P4,P5} for the first time the shadow image of the supermassive black hole in the center of M8$7^{*}$ galaxy, which greatly stimulates our enthusiasm for the research of black hole shadows. The essence of shadows is that the specific photons around a black hole collapse inward to form a dark area observed by a distant observer, which shows that the shadow appears to be a dynamical phenomenon of black holes. Through black hole shadows, one not only obtains~\cite{P6,P7,P8,P10,PPQQ1} the mass, spin, charge, and other information of black holes, but also knows~\cite{P11,PHXW,P12,P13} the distribution of matter around black holes. More importantly, the relevant research opens a new window for us to study the strong gravitational region near black hole horizons. So far, there have been a large number of articles on black hole shadows under Einstein gravity and modified gravity theories, see, for instance, some literature~\cite{P44,P45,TTHHH1,P46,P47,P48,P50,P51,P52,P53,PX90,P54,P55,P56,P57,P58,JM1,JM2,P59,P71,P60,P61,PDD1,PDD3,PDD4,PDD5,PDD6,PDD7,PDD9,PDD10,PDD11, TTDD71,TTDD72,TTDD73, PDD12,PDD13,PDD23,PDD24,PDD25,PPQQ2,PTT1,POO724}.

A shadow is an observable quantity which is potentially related to black hole thermodynamics. As is known, the black hole thermodynamics~\cite{P14,P15,P16,PDD26,PDD27} is an important subject of black hole physics, especially when combined with AdS spacetimes, the black hole thermodynamics in extended phase spaces has achieved great progress, {\em e.g.}, the van der Waals-like phase transition occurs~\cite{P17,PDD18,PDD19,PDD21,PDD22} in the Reissner-Nordstr$\mathrm{ \ddot{o}}$m AdS black hole, the reentrant phase transition happens~\cite{P18} in high-dimensional rotating black holes, and so on. Now black holes are not only regarded as a strong gravitational system, but also as a thermodynamic one, so it is necessary to explore the relationship between its dynamics and thermodynamics. Recently, the relationships between dynamic characteristic quantities and thermodynamic ones have been studied~\cite{P19,P20,P21,P22,P23,P24,P25,P26,P27,P28}.
Some results show~\cite{P19,P20} that the unstable circular orbital motion can connect thermodynamic phase transitions of black holes. For a static spherically symmetric black hole, its timelike or lightlike circular orbital motion of a test particle and its shadow indeed have a close connection~\cite{P22,P27,P28} to its thermodynamic phase transition.  Moreover,  for a rotating black hole  with  the cosmological constant its shadow radius can also reflect~\cite{P27} its thermodynamic phase transition.

The Ruppeiner thermodynamic geometry~\cite{P29,P30,P31,P32} is a useful tool to study black hole thermodynamics. It provides~\cite{P33,P34,P35,P36,P37,P38,P39,P40,P41,P42,P43,PDD14,PDD15,PDD16,PDD17} helpful information about microstructure of black holes. Its most important physical quantity is the Ruppeiner thermodynamic scalar curvature. It has been shown~\cite{P30,P32,P38,P41,P42,P43} that this scalar curvature can take a positive, or a negative, or a vanishing value which corresponds to a repulsive, or an attractive, or no interaction among black hole molecules, respectively. As is known,  the black hole microstructure represents an important aspect of black hole thermodynamics. Therefore, it is necessary to associate a shadow radius with the Ruppeiner thermodynamic scalar curvature, so as to connect such an observable to black hole microstructure.

We emphasize our main  motivation:
Considering that the thermodynamic phase transition and microstructure of black holes are not detectable directly,  we want to connect certain observables  that describe  the characteristics of shadow deformations of rotating black holes  to the two aspects of black hole thermodynamics, thus taking a glimpse of the inside of a black hole from the outside.
Our investigations may provide some helps for understanding black hole thermodynamics via black hole shadows.

The paper is organized as follows. In Sec. 2, we investigate the relationship between shadow  radius and microstructure for a  general  static spherically symmetric black hole. Then, we  take Reissner-Nordstr\"om (AdS) black hole as an example to do the concrete analysis in Sec. 3.  We study the relationship between shadow and  thermodynamics in the aspects of phase transition and microstructure  for the  Kerr (AdS)  black hole in Sec. 4.  We  give  in Sec. 5 the  constraints to the relaxation time of  black hole perturbation based on shadow data. Finally, we make a summary in Sec. 6. We use the units $c=G=k_{\rm B}=\hbar=1$ and the sign convention $(-,+,+,+)$ throughout this paper.

\section{Relationship between shadow  radius  and  microstructure  for a  static spherically symmetric black hole}

A  general static spherically symmetric black hole can be described by the line element,
\begin{equation}
\label{1}
ds^{2}=-f(r)dt^{2}+\frac{dr^{2}}{g(r)}+r^{2}\left(d\theta^{2}+\sin^{2}\theta d\varphi^{2}\right),
\end{equation}
where $f(r)$ and $g(r)$ are functions of radial coordinate $r$.  The Hamilton-Jacobi equation can be used to separate  the null geodesic equations in the spacetime of static spherically symmetric black holes and the general expression of shadows can be derived for this class of black holes. The Hamilton-Jacobi equation takes~\cite{PSCH} the form,
\begin{equation}
\label{2}
\frac{\partial {\cal S}}{\partial \lambda }+{\cal H}=0, \qquad    {\cal H}=\frac{1}{2}g_{\mu \nu }p^{\mu }p^{\nu },
\end{equation}
where $\lambda$ is the affine parameter of null geodesics,  ${\cal S}$ the Jacobi action, ${\cal H}$ the  Hamiltonian, and the  momentum $p^{\mu }$ is defined by
\begin{equation}
\label{3}
p_{\mu }\equiv \frac{\partial {\cal S}}{\partial x^{\mu }}=g_{\mu \nu }\frac{\mathrm{d} x^{\nu }}{\mathrm{d}\lambda}.
\end{equation}

Due to the spherical symmetry of Eq.~(\ref{1}), without loss of generality, one can consider the photons moving on the equatorial plane with $\theta =\frac{\pi }{2}$.  So the the Jacobi action ${\cal S}$ can be decomposed into the following form,
\begin{equation}
\label{4}
{\cal S}=\frac{1}{2}m^{2}\lambda -Et+L\varphi +S_{r}(r),
\end{equation}
where $m$ is the mass of moving particles around a static spherically symmetric black hole ($m=0$ for photons), $E$ the energy of photons, $L$ the angular momentum of photons, and $S_{r}(r)$ a function of coordinate $r$. By substituting Eqs.~(\ref{1}), (\ref{3}) and (\ref{4}) into Eq.~(\ref{2}), one can get the following three equations describing the motion of photons on the equatorial plane,
\begin{eqnarray}
\frac{\mathrm{d} t}{\mathrm{d} \lambda }& =& \frac{E}{f(r)},\label{5} \\
\frac{\mathrm{d} r}{\mathrm{d} \lambda }&=& \pm \frac{\sqrt{g(r)}}{\sqrt{f(r)}}\frac{\sqrt{E^{2}r^{4}-L^{2}r^{2}f(r)}}{r^{2}},\label{6} \\
\frac{\mathrm{d} \varphi }{\mathrm{d} \lambda }&=& \frac{L}{r^{2} }.\label{7}
\end{eqnarray}
Combining Eq.~(\ref{6}) with Eq.~(\ref{7}), one has
\begin{equation}
\label{8}
\frac{\mathrm{d} r}{\mathrm{d}\varphi }= \frac{\frac{\mathrm{d} r }{\mathrm{d} \lambda }}{\frac{\mathrm{d \varphi} }{\mathrm{d} \lambda }}=\pm r\sqrt{g(r)\left(\frac{r^{2}E^{2}}{L^{2}f(r)} -1\right )}.
\end{equation}
Again considering $\left.\frac{\mathrm{d} r}{\mathrm{d}\varphi }\right|_{r=\xi}=0$ at the turning point of photon orbits, one can rewrite Eq.~(\ref{8}) as
\begin{equation}
\label{9}
\frac{\mathrm{d} r}{\mathrm{d}\varphi }= \pm r\sqrt{g(r)\left(\frac{r^{2}f(\xi )}{\xi^{2}f(r)} -1\right )}.
\end{equation}
In addition, according to the equation satisfied by the effective potential $V(r)$,
\begin{equation}
\label{10}
\left (\frac{\mathrm{d} r}{\mathrm{d} \lambda }\right)^{2}+V(r)=0,
\end{equation}
one obtains
\begin{equation}
\label{11}
V(r)=g(r)\left(\frac{L^{2}}{r^{2}}-\frac{E^{2}}{f(r)}\right).
\end{equation}
Using the effective potential, one can determine~\cite{PSCH} the critical orbit radius of photons, {\em i.e.} the photosphere radius in a static spherically symmetric spacetime,
\begin{equation}
\label{12}
V(r)=0, \qquad \frac{\mathrm{d} V(r)}{\mathrm{d} r}=0,  \qquad  \frac{d^2 V(r)}{d r^2}<0.
\end{equation}
By substituting  Eq.~(\ref{11}) into Eq.~(\ref{12}), one can obtain the following relationships,
\begin{equation}
\label{13}
\frac{L^{2}}{E^{2}}=\frac{r^{2}_{\rm p}}{f(r_{\rm p})},
\end{equation}
\begin{equation}
\label{14}
r \frac{\mathrm{d} f(r)}{\mathrm{d} r}-2f(r)\bigg|_{r=r_{\rm p}}=0,
\end{equation}
where $r_{\rm p} $ is the radius of photospheres.

Considering that the light emitted from  a static observer at  position $r_{0}$  transmits into the past with an angle $\vartheta $ relative to the radial direction, one obtains~\cite{P53,PX90,P27}
\begin{equation}
\label{15}
\cot\vartheta =\frac{\sqrt{g_{rr}}}{\sqrt{g_{\varphi \varphi }}}\frac{\mathrm{d} r}{\mathrm{d}\varphi }\bigg|_{r=r_{0} }=\pm \sqrt{\frac{{r_{0}}^{2}f(\xi )}{\xi^{2}f(r_{0})} -1}.
\end{equation}
When the relation $\sin^{2}\vartheta =\frac{1}{1+\cot^{2}\vartheta }$ is used, the above equation can be written as
\begin{equation}
\label{16}
\sin^{2}\vartheta =\frac{{\xi }^{2}f(r_{0})}{{r_{0}}^{2}f(\xi)}.
\end{equation}
Therefore, the shadow radius of black holes observed by a static observer at position $r_{0}$ can be expressed~\cite{P13,P53,PX90,P27} by
\begin{equation}
\label{17}
r_{\rm sh}\equiv r_{0}\sin\vartheta =\left.\xi \sqrt{\frac{f(r_{0})}{f(\xi )}}\right|_{\xi \rightarrow r_{\rm p}}.
\end{equation}

In the Ruppeiner thermodynamic geometry, the most important physical quantity that describes the microstructure of black holes is the Ruppeiner thermodynamic scalar curvature $R$~\cite{P29,P30}. If the entropy representation is chosen~\cite{P64,P65,P72}, this scalar curvature can be expressed by a function of event horizon radius and the other parameters,\footnote{The other parameters usually include charge, angular momentum, pressure, etc., which depends on models. As we are going to analyze only the relation between event horizon radius and shadow radius, we write the event horizon radius explicitly but hide the other parameters. Moreover, ``..." means the other variables in the phase space, see, for instance, it means thermo-charge in Eq.~(\ref{33}) for the Reissner-Nordstr\"om (AdS) black hole and angular momentum in Eq.~(\ref{52}) for the Kerr (AdS) black hole.}
\begin{equation}
\label{18}
R(S, ...)=R(S(r_{\rm H}), ...),
\end{equation}
which leads to
\begin{equation}
\label{ZJ1}
\frac{\partial R}{\partial  r_{\rm H}}=\frac{\partial R}{\partial r_{\rm sh}}\frac{\mathrm{d} r_{\rm sh}}{\mathrm{d}r_{\rm H}}.
\end{equation}
As a result,  according to the relation~\cite{P22,P27},
\begin{equation}
\label{19}
 \frac{\mathrm{d} r_{\rm sh} }{\mathrm{d} r_{\rm H}}>0,
\end{equation}
we know that the conditions
\begin{equation}
\label{21}
\frac{\partial R}{\partial r_{\rm H}}>0, \qquad   \frac{\partial R}{\partial  r_{\rm H}}=0, \qquad   \frac{\partial R}{\partial  r_{\rm H}}<0,
\end{equation}
 can be converted to
\begin{equation}
\label{22}
\frac{\partial R}{\partial r_{\rm sh}}>0, \qquad   \frac{\partial R}{\partial r_{\rm sh}}=0, \qquad   \frac{\partial R}{\partial r_{\rm sh}}<0.
\end{equation}
We thus establish the relationship between shadow  radius and microstructure for a  general  static spherically symmetric black hole. Next, we take the  Reissner-Nordstr$\mathrm{\ddot{o}}$m  (RN) AdS  black hole  as an example to  analyze  the  connection  between  Ruppeiner thermodynamic  scalar curvature  and shadow  radius.

\section{Relationship between shadow and   microstructure  for  RN (AdS) black holes}

The line element of the RN AdS black hole is given  by~\cite{PDD26,P17}
\begin{equation}
\label{23}
ds^{2}=-f(r)dt^{2}+\frac{dr^{2}}{f(r)}+r^{2}\left(d\theta^{2}+\sin^{2}\theta d\varphi^{2}\right),
\end{equation}
with
\begin{equation}
\label{24}
f(r)=1-\frac{2M}{r}+\frac{q^{2}}{r^{2}}-\frac{\Lambda r^{2}}{3}.
\end{equation}
Here $M$ is the black hole mass, $q$ the charge, $\Lambda$ the negative cosmological constant. In the extended phase space including the cosmological constant,  some thermodynamic quantities  for the RN AdS black hole are as follows~\cite{PDD26,P17,PDD15,PDD16}:
\begin{eqnarray}
M=H&=&\frac{r_{\rm H}}{2}+\frac{Q}{2r_{\rm H}}+\frac{4\pi P}{3}r^{3}_{\rm H}\nonumber \\
&=&\frac{S (8 P S+3)+3 \pi  Q}{6 \sqrt{\pi  S}},\label{25}\\
T&=&\frac{1}{4\pi r_{\rm H}}-\frac{Q}{4\pi r^{3}_{\rm H} }+2Pr_{\rm H}\nonumber \\
&=&\frac{8 P S^2-\pi  Q+S}{4 S \sqrt{\pi  S}},\label{26}\\
S&=&\pi r_{\rm H}^{2},\label{27}\\
V&=&\frac{4\pi r^{3}_{\rm H} }{3},\label{28}\\
P&=&-\frac{\Lambda }{8\pi },\label{29}
\end{eqnarray}
where $M$  is  regarded as the black hole enthalpy  $H$,  $T$  the Hawking temperature, $S$  the entropy,  $V$  the thermo-volume, $P$  the pressure,\footnote{If $\Lambda=0$ or $P=0$, the RN AdS black hole reduces to the RN black hole.}  $ r_{\rm H}$  the event horizon radius,  and the thermo-charge  $Q\equiv q^{2}$. In addition, the first law of thermodynamics and the Smarr relation take~\cite{PDD15,PDD16} the forms,
\begin{eqnarray}
dM&=&TdS+VdP+\Psi dQ,\label{30}\\
M&=&2(TS-PV+\Psi Q),\label{31}
\end{eqnarray}
where  the thermal-potential $\Psi \equiv \left(\frac{\partial M}{\partial Q}\right)_{S,P}=\frac{1}{2r_{\rm H}}$.

The Ruppeiner  line element of the RN AdS black hole can be written in the phase space $\left \{S, Q\right\}$  as follows~\cite{PDD15,PDD16}:
\begin{equation}
\label{32}
ds^{2}_{\rm R}=\frac{1}{T}\left( \frac{\partial T}{\partial S}\right)_{Q}dS^{2}+\frac{2}{T}\left( \frac{\partial T}{\partial Q}\right)_{S}dSdQ,
\end{equation}
and then the thermodynamic curvature $R_{\rm SQ}$ can be obtained~\cite{PDD15,PDD16} from Eqs.~(\ref{26}),  (\ref{27})  and  (\ref{32}),
\begin{equation}
\label{33}
R_{\rm SQ}=\frac{16 P S+1}{8 P S^2-\pi  Q+S}.
\end{equation}

By substituting Eqs.~(\ref{24}) and (\ref{25}) into Eqs.~(\ref{14}) and (\ref{17}),  we calculate the shadow radius $r_{\rm sh}$ of the RN AdS black hole  observed by a static observer at position $r_{0}$,
\begin{eqnarray}
\label{34}
r_{\rm sh}&=&\frac{\left(\sqrt{B+8 Q}+\sqrt{B}\right)^2 \sqrt{r_0 \left(-2 \sqrt{B+8 Q}+8 \pi  P r_0^3+3 r_0\right)+3 Q}}{2 \sqrt{2} r_0 \sqrt{8 \pi  B^2 P+8 \pi  B P \left(\sqrt{B (B+8 Q)}+8 Q\right)+\left(\sqrt{B (B+8 Q)}+2 Q\right) (32 \pi  P Q+1)+B}}  \nonumber \\
&=&r_{\rm sh}(r_{\rm H}, Q, P, r_0),
\end{eqnarray}
with
\begin{equation}
\label{35}
B\equiv \frac{\left(8 \pi  P r_{\rm H}^4+3 r_{\rm H}^2+3 Q\right){}^2-32 Q r_{\rm H}^2}{4 r_{\rm H}^2}.
\end{equation}
Again substituting Eq.~(\ref{27}) into Eq.~(\ref{33}), we derive the Ruppeiner thermodynamic scalar curvature  $R_{\rm SQ}$ as a function of $r_{\rm H}$, $Q$  and  $P$,
\begin{equation}
\label{36}
R_{\rm SQ}=R_{\rm SQ}(r_{\rm H}, Q, P)=\frac{16 \pi  P r_{\rm H}^2+1}{8 \pi ^2 P r_{\rm H}^4+\pi  r_{\rm H}^2-\pi  Q}.
\end{equation}

According to Eqs.~(\ref{34}) and (\ref{36}), we deduce that there is a correspondence between  the Ruppeiner thermodynamic scalar curvature $R_{\rm SQ}$ and  the shadow radius $r_{\rm sh}$ for the RN (AdS) black hole  if charge $Q$ and  pressure $P$ are fixed. We cannot solve $R_{\rm SQ}(r_{\rm sh})$ analytically but express it as follows:
\begin{equation}
\label{37}
R_{\rm SQ}=\left.R_{\rm SQ}(r_{\rm sh})\right|_{Q, \,P}.
\end{equation}
Based on Eq.~(\ref{37}), we depict the relation  of the Ruppeiner thermodynamic scalar curvature $R_{\rm SQ}$ with respect to  the shadow radius $r_{\rm sh}$  at constant pressure $P$ and constant charge $Q$ in Fig. 1,  where  the red  curve  $R_{\rm SQ}-r_{\rm H}$  is  attached  as a comparison.  By comparing the blue curve with the red one, we can clearly see that they have similar profiles, which indicates that  the shadow radius $r_{\rm sh}$ connects to the  black hole  microstructure as the event horizon radius $r_{\rm H}$ does in Eq.~(\ref{36}). Therefore, the shadow radius is indeed a good observable that has a close connection to the black hole microstructure  for a  static spherically symmetric black hole.

\begin{figure}
		\centering
		\begin{minipage}{.5\textwidth}
			\centering
			\includegraphics[width=85mm]{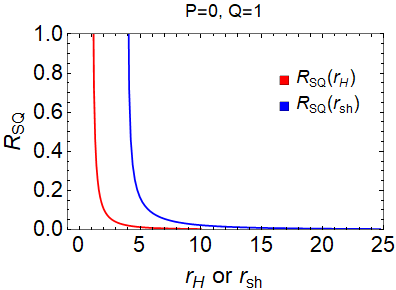}
		\end{minipage}%
		\begin{minipage}{.5\textwidth}
			\centering
			\includegraphics[width=85mm]{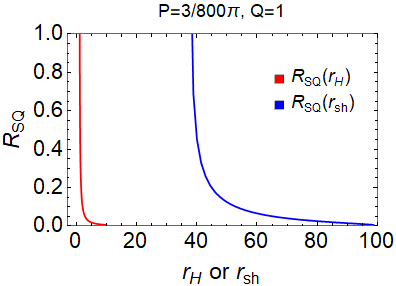}
		\end{minipage}

		\caption*{Fig. 1. Ruppeiner thermodynamic scalar curvature  $R_{\rm SQ}$  with respect to  the event horizon radius  $r_{\rm H}$ and  to the shadow radius $r_{\rm sh}$  at constant pressure $P$ and constant charge $Q$. Here we set $r_{\rm 0}=100$. The left graph corresponds to the RN black hole and the right one to the RN AdS black hole.}
\label{figure1}
\end{figure}

\section{Relationship between  shadow  and  thermodynamics in the aspects of phase transition and microstructure for  Kerr (AdS)  black holes}

\subsection{Hawking temperature and Ruppeiner thermodynamic scalar curvature with respect to event horizon radius}
We take  the Kerr (AdS)  black hole  as a  representative of  rotating  black holes and study its relationship  between shadow and thermodynamics in the aspects of  phase transition and  microstructure.

The line element of the Kerr AdS black hole takes~\cite{PDD28} the form,
\begin{eqnarray}
\label{38}
ds^{2}&=&-\frac{(\Delta _{r}-\Delta _{\theta }a^{2}\sin^{2}\theta)}{\Sigma }dt^{2}-\frac{2a\sin^{2}\theta}{\Xi \Sigma }[(r^{2}+a^{2})\Delta _{\theta }-\Delta _{r}]dtd\varphi +\frac{\Sigma }{\Delta _{r}}dr^{2}+\frac{\Sigma }{\Delta _{\theta }}d\theta ^{2}\nonumber \\
& & +\frac{\sin^{2}\theta}{\Xi ^{2}\Sigma }[(r^{2}+a^{2})^2\Delta _{\theta }-\Delta _{r}a^{2}\sin^{2}\theta]d\varphi^{2},
\end{eqnarray}
with
\begin{eqnarray}
\Sigma &=&r^{2}+a^{2}\cos^{2}\theta,\label{39}\\
\Delta _{r}&=&\left(a^2+r^2\right) \left(1-\frac{\Lambda  r^2}{3}\right)-2 m r,\label{40}\\
\Delta _{\theta }&=&1+\frac{\Lambda }{3}a^2\cos^2\theta,\label{41}\\
\Xi &=&1+\frac{\Lambda }{3}a^2,\label{42}
\end{eqnarray}
where $m$ and $a$ are parameters related to the mass and rotation of the  black hole, respectively.  In the extended phase space including the cosmological constant  $\Lambda$,  some thermodynamic quantities  for the  Kerr AdS black hole read~\cite{PDD27,PDD18,PDD19,PDD21,PDD22,P18} as follows:
\begin{eqnarray}
\label{43}
M&=&H=\sqrt{\frac{(8 P S+3) \left[12 \pi ^2 J^2+S^2 (8 P S+3)\right]}{36 \pi  S}},\\
\label{44}
T&=&\frac{S^2 \left(64 P^2 S^2+32 P S+3\right)-12 \pi ^2 J^2}{4 \sqrt{S^3 (8 P S+3) \left[12 \pi ^3 J^2+\pi  S^2 (8 P S+3)\right]}},\\
\label{45}
\Omega& =&\frac{2 \pi ^{3/2} J \sqrt{8 P S+3}}{\sqrt{S \left[12 \pi ^2 J^2+S^2 (8 P S+3)\right]}},\\
\label{46}
V&=&\frac{4 \sqrt{S} \left[6 \pi ^2 J^2+S^2 (8 P S+3)\right]}{3 \sqrt{8 P S+3} \sqrt{12 \pi ^3 J^2+\pi  S^2 (8 P S+3)}},\\
\label{47}
S&=&\frac{\pi  \left(a^2+r_{\rm H}^2\right)}{\Xi },\\
\label{48}
P&=&-\frac{\Lambda }{8\pi },
\end{eqnarray}
where the ADM mass $M= m/\Xi^{2}$  is  regarded as the black hole enthalpy  $H$,  $T$  the Hawking temperature, $S$  the entropy,  $V$  the thermo-volume, $J=Ma$  the angular momentum,  $\Omega$   the angular velocity, $P$   the pressure,  and  $r_{\rm H}$  the event horizon radius. In addition, the first law of thermodynamics and the Smarr relation are~\cite{PDD27,PDD18,PDD19,PDD21,PDD22,P18}
\begin{equation}
\label{49}
dM=TdS+VdP+\Omega dJ,
\end{equation}
\begin{equation}
\label{50}
M=2(TS-PV+\Omega J).
\end{equation}

The Ruppeiner  line element describing this black hole in the  phase space $\left \{ S,J \right \}$  has~\cite{PDD14} the form,
\begin{equation}
\label{51}
ds^{2}_{\rm R}=\frac{1}{T}\left( \frac{\partial T}{\partial S}\right)_{J}dS^{2}+\frac{2}{T}\left( \frac{\partial T}{\partial J}\right)_{S}dSdJ+\frac{1}{T}\left( \frac{\partial \Omega}{\partial J}\right)_{S}dJ^{2}.
\end{equation}
Using Eqs.~(\ref{44}),  (\ref{45}),   and   (\ref{51}),  one can write~\cite{P32,PDD14} the thermodynamic curvature,
\begin{equation}
\label{52}
R_{\rm SJ}=-\frac{1}{\sqrt{g}}\left[\frac{\partial }{\partial S}\left(\frac{g_{12}}{g_{11}\sqrt{g}}\frac{\partial g_{11}}{\partial J}-\frac{1}{\sqrt{g}}\frac{\partial g_{22} }{\partial S}\right)+\frac{\partial }{\partial J}\left(\frac{2}{\sqrt{g}}\frac{\partial g_{12}}{\partial S}-\frac{1}{\sqrt{g}}\frac{\partial g_{11}}{\partial J}-\frac{g_{12}}{g_{11}\sqrt{g}}\frac{\partial g_{11}}{\partial S}\right)\right],
\end{equation}
where
\begin{eqnarray}
\label{53}
g_{11}&=& -\frac{3 S (4 P S+1)}{12 \pi ^2 J^2+S^2 (8 P S+3)}+\frac{2 S \left[16 P S (8 P S+3)+3\right]}{S^2 (8 P S+1) (8 P S+3)-12 \pi ^2 J^2}-\frac{4 P}{8 P S+3}-\frac{3}{2 S},\\
\label{54}
g_{12}&=& g_{21}=12 \pi ^2 J \left[\frac{2}{12 \pi ^2 J^2-S^2 (8 P S+1) (8 P S+3)}-\frac{1}{12 \pi ^2 J^2+S^2 (8 P S+3)}\right],\\
\label{55}
g_{22}&=&\frac{8 \pi ^2 S^5 (8 P S+3)^{5/2}}{S^2 \left[12 \pi ^2 J^2+S^2 (8 P S+3)\right] \left[S^2 (8 P S+1) (8 P S+3)-12 \pi ^2 J^2\right]\sqrt{ (8 P S+3) }},\\
\label{56}
g&=&g_{11}g_{22}-g_{12}^{2}.
\end{eqnarray}
Substituting Eq.~(\ref{47}) into Eq.~(\ref{44}), and Eqs.~(\ref{43}) and (\ref{47}) into Eq.~(\ref{52}),  we know that the Hawking temperature $T$ and the Ruppeiner thermodynamic scalar curvature  $R_{\rm SJ}$ are functions of $r_{\rm H}$, $J$, and $P$,
\begin{eqnarray}
\label{565}
T&=&T(r_{\rm H}, J, P),\\
\label{57}
R_{\rm SJ}&=&R_{\rm SJ}(r_{\rm H}, J, P).
\end{eqnarray}

\subsection{Hawking temperature and Ruppeiner thermodynamic scalar curvature with respect to shadow radius, or deformation parameters, or circularity deviation}
The null geodesic equations of the photons around the Kerr AdS  black hole can be expressed~\cite{PDD9,PDD13} as
\begin{eqnarray}
\label{58}
\frac{\mathrm{d} t}{\mathrm{d}\lambda }&=&\frac{\left(a^2 \sin ^2\theta +\Sigma \right) \left(a^2 E \sin ^2\theta -a L \Xi +\Sigma  E\right)}{\Sigma  \Delta _r}+\frac{a \left(L \Xi -a E \sin ^2\theta \right)}{\Sigma  \Delta _{\theta }},\\
\label{59}
\frac{\mathrm{d} r}{\mathrm{d} \lambda }&=&\frac{\sqrt{\left[E \left(a^2 \sin ^2\theta +\Sigma \right)-a L \Xi \right]^2-\Delta _r \left[(L \Xi -a E)^2+\text{K}\right]}}{\Sigma }\equiv \frac{\sqrt{U(r)}}{\Sigma },\\
\label{60}
\frac{\mathrm{d} \theta }{\mathrm{d} \lambda }&=&\frac{\sqrt{\Delta _{\theta } \left[(L \Xi -a E)^2+\text{K}\right]-(L \Xi  \csc \theta -a E \sin \theta )^2}}{\Sigma }\equiv \frac{\sqrt{\Theta (\theta )}}{\Sigma },\\
\label{61}
\frac{\mathrm{d} \varphi }{\mathrm{d} \lambda }&=&\frac{a\Xi  \left[E \left(a^2 \sin ^2\theta +\Sigma \right)-a L \Xi \right]}{ \Sigma  \Delta _r}+\frac{L \Xi^{2} -a\Xi E \sin ^2\theta }{ \Sigma  \Delta _{\theta } \sin ^2\theta },
\end{eqnarray}
where  $L$ and $E$ are the angular momentum and energy of photons,  respectively,  $\lambda$  is the affine parameter, $K$  is the  Carter constant, and $U(r)$  and  $\Theta (\theta )$  are non-negative definite functions of $r$ and $\theta$, respectively.

The radius of the unstable  orbit of  photons $r_{\rm p}$ can be solved~\cite{PDD9,PDD13} from the following conditions:
\begin{equation}
\label{62}
 U(r_{\rm p})=0,   \qquad      \left.\frac{\mathrm{d} U(r)}{\mathrm{d} r}\right|_{r=r_{\rm p}}=0,
\end{equation}
where the definition of $U(r)$ is given in Eq.~(\ref{59}). The two conditions can be expressed explicitly as~\cite{PDD9,PDD13}
\begin{eqnarray}
\label{63}
L_{E}&\equiv& \frac{L}{E}=\frac{3 \left[a^4 \Lambda  r_{\rm p}+a^2 \left(3 m+\Lambda  r_{\rm p}^3+3 r_{\rm p}\right)+3r_{\rm p}^2 \left(r_{\rm p}-3 m\right)\right]}{a \left(a^2 \Lambda +3\right) \left[r_{\rm p} \left(a^2 \Lambda +2 \Lambda  r_{\rm p}^2-3\right)+3 m\right]},\\
\label{64}
K_{E}&\equiv& \frac{K}{E^{2}}=-\frac{r_{\rm p}^3 \left\{a^4 \Lambda ^2 r_{\rm p}^3+6 a^2 \left[3 m \left(\Lambda r_{\rm p}^2-2\right)+\Lambda  r_{\rm p}^3\right]+9 r_{\rm p} \left(r_{\rm p}-3 m\right)^2\right\}}{a^2 \left[r_{\rm p} \left(a^2 \Lambda +2 \Lambda  r_{\rm p}^2-3\right)+3 m\right]^2}.
\end{eqnarray}
After substituting  Eqs.~(\ref{63}) and (\ref{64}) into   Eq.~(\ref{60}), one can determine the value range of $r_{\rm p}$,
\begin{equation}
\label{65}
r_{\rm p}\in [r_{\rm p1},r_{\rm p2}].
\end{equation}

To describe the contour of the black hole shadow, we consider an  observer  located  in the zero-angular-momentum reference  frame~\cite{PDD29}  as follows:
\begin{eqnarray}
\label{66}
\hat{e}_{(t)}&=&\sqrt{\frac{g_{\varphi \varphi }}{g_{t\varphi }^{2}-g_{tt}g_{\varphi \varphi }}}\left(\partial _{t}-\frac{g_{t\varphi }}{g_{\varphi \varphi } }\partial_{\varphi }\right),\\
\label{67}
\hat{e}_{(r)}&=&\frac{1}{\sqrt{g_{rr}}}\partial _{r},\\
\label{68}
\hat{e}_{(\theta )}&=&\frac{1}{\sqrt{g_{\theta \theta }}}\partial _{\theta },\\
\label{69}
\hat{e}_{(\varphi )}&=&\frac{1}{\sqrt{g_{\varphi \varphi }}}\partial _{\varphi },
\end{eqnarray}
where the timelike vector  $\hat{e}_{(t)}$    and   the  spacelike vectors $\hat{e}_{(r)}$, $\hat{e}_{(\theta )}$, $\hat{e}_{(\varphi )}$  are  normalized orthogonally to each other.  For null geodesics, the four momentum  $p^{\mu }$ of photons can be projected~\cite{PDD13} onto the four bases of the observer's frame, i.e.
\begin{eqnarray}
\label{70}
p^{(t)}&=&-p_{\mu }\hat{e}^{\mu }_{(t)},\\
\label{71}
p^{(i)}&=&p_{\mu }\hat{e}^{\mu }_{(i)},
\end{eqnarray}
where the index $i$ means the space coordinates, $r,\theta ,\varphi$.

The observation angles  $(\alpha, \beta )$  are  introduced~\cite{PDD30} by the formulas,
\begin{eqnarray}
\label{72}
p^{(r)}&=&p^{(t)}\cos\alpha \cos\beta,\\
\label{73}
p^{(\theta )}&=&p^{(t)}\sin\alpha,\\
\label{74}
p^{(\varphi )}&=&p^{(t)}\cos\alpha \sin\beta.
\end{eqnarray}
Combining Eqs.~(\ref{58})-(\ref{74}), one can obtain~\cite{P27,PDD13} the relations satisfied by the observation angles,\footnote{The notations: $\hat{e}^{t}_{(t)}= \sqrt{\frac{g_{\varphi \varphi }}{g_{t\varphi }^{2}-g_{tt}g_{\varphi \varphi }}}$ and $\hat{e}^{\varphi }_{(t)}=-\frac{g_{t\varphi }}{g_{\varphi \varphi } }\sqrt{\frac{g_{\varphi \varphi }}{g_{t\varphi}^{2}-g_{tt}g_{\varphi \varphi }}}$, see Eq.~(\ref{66}).}
\begin{eqnarray}
\label{75}
\sin\alpha&=&\frac{p^{(\theta )}}{p^{(t)}}=\left.\pm \frac{1}{\hat{e}^{t}_{(t)}-L_{E}\hat{e}^{\varphi  }_{(t)}}\sqrt{\frac{{\Delta _{\theta } \left[(L_{E} \Xi -a)^2+K_{E}\right]-(L_{E} \Xi  \csc \theta -a \sin \theta )^2}}{\Sigma  \Delta _{\theta }}}\right|_{(r_{0}, \theta _{0})},\\
\label{76}
\tan\beta &=&\left.\frac{p^{(\varphi )}}{p^{(r)}}=L_{E}{\sqrt\frac{\Sigma\Delta _r} {{g_{\varphi \varphi}\left[\left(a^2 \sin ^2\theta -a L_{E} \Xi +\Sigma \right)^2-\Delta _r \left((L_{E} \Xi -a)^2+K_{E}\right)\right]}}}\right|_{(r_{0}, \theta _{0})},
\end{eqnarray}
where  $r_{0}$  is  the distance between the observer and the black hole, and $\theta _{0}$ is the inclination angle between the observer's sight line and the black hole's rotation axis.  In addition, one can determine the apparent position  on the sky plane of the observer by introducing the following Cartesian coordinates~\cite{PSCH,PDD12,PDD13},
\begin{eqnarray}
\label{77}
x&=&-r_{0}\cos\alpha \sin\beta,\\
\label{78}
y&=&r_{0}\sin\alpha.
\end{eqnarray}

According to the schematic diagram in Fig. 2, the shadow radius $r_{\rm sh}$, the deformation parameters,  $\delta _{s}$ and $k_{s}$, and  the circularity deviation  $\Delta C$  for a  rotating black hole  can be defined~\cite{PDD31,PDD32,PDD33,PDD34} as follows:
\begin{eqnarray}
\label{79}
r_{\rm sh}&\equiv&\frac{1}{2\pi } \int_{0}^{2\pi }l(\phi )d\phi,\\
\label{80}
\delta _{s}&\equiv& \frac{2r_{\rm sh}-(x_{\rm max}-x_{\rm min})}{r_{\rm sh}},\\
\label{81}
k_{s}&\equiv& \frac{2y_{h}}{x_{\rm max}-x_{\rm min}},\\
\label{82}
\Delta C&\equiv& 2\sqrt{\frac{1}{2\pi }\int_{0}^{2\pi }(l(\phi )-r_{\rm sh})^{2}d\phi },
\end{eqnarray}
where
\begin{eqnarray}
\label{83}
l(\phi )&\equiv&\sqrt{(x-x_{c})^{2}+(y-y_{c})^{2}},\\
\label{84}
 x_{c}&\equiv& \frac{|x_{\rm min}+x_{\rm max}|}{2}, \\
 y_{c}&\equiv&0,\\
\label{85}
\tan\phi &\equiv&\frac{y-y_{c}}{x-x_{c}}.
\end{eqnarray}

\begin{figure}
\centering
\begin{minipage}[t]{0.6\linewidth}
\centering
\includegraphics[width=100mm]{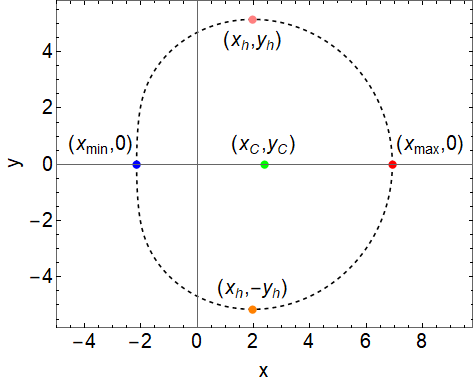}
\caption*{Figure 2. A schematic diagram for a  rotating  black hole shadow.}
\label{fig2}
\end{minipage}
\end{figure}

Since the  Kerr AdS black hole is completely determined  by  only three parameters ($m$, $J$, $P$), we obtain the results,
\begin{eqnarray}
\label{86}
r_{\rm sh}&=&r_{\rm sh}(r_{\rm H}, J, P),\\
\label{87}
\delta _{s}&=&\delta _{s}(r_{\rm H}, J, P),\\
\label{88}
k_{s}&=&k_{s}(r_{\rm H}, J, P),\\
\label{89}
\Delta C&=&\Delta C(r_{\rm H}, J, P).
\end{eqnarray}
This means that the shadow radius, deformation parameters, and circularity deviation can be expressed as functions of the event horizon radius, angular momentum, and pressure.

According to Eqs.~(\ref{565}) and (\ref{57}) and Eqs.~(\ref{86})-(\ref{89}), we can express both  the Hawking temperature  $T$ and the Ruppeiner thermodynamic scalar curvature $R_{\rm SJ}$ as a one-variable function of  the shadow radius $r_{\rm sh}$, or the deformation parameter  $\delta _{s}$ or $k_{s}$, or  the circularity deviation  $\Delta C$  for the Kerr (AdS) black hole  if  the  angular momentum  $J$ and  pressure $P$ are fixed, 
\begin{equation}
\label{92}
T=\left. T(r_{\rm sh})\right|_{J, \,P},   \qquad     R_{\rm SJ}=\left.R_{\rm SJ}(r_{\rm sh})\right|_{J, \,P},
\end{equation}
\begin{equation}
\label{93}
T=\left.T(\delta _{s})\right|_{J, \,P},   \qquad     R_{\rm SJ}=\left.R_{\rm SJ}(\delta _{s})\right|_{J, \,P},
\end{equation}
\begin{equation}
\label{94}
T=\left.T(k_{s})\right|_{J, \,P},   \qquad     R_{\rm SJ}=\left.R_{\rm SJ}(k_{s})\right|_{J, \,P},
\end{equation}
\begin{equation}
\label{95}
T=\left.T(\Delta C)\right|_{J, \,P},     \qquad    R_{\rm SJ}=\left.R_{\rm SJ}(\Delta C)\right|_{J, \,P}.
\end{equation}

Based on Eqs.~(\ref{86})-(\ref{89}),  we depict in Figs. 3 and 4 for the Kerr  and Kerr AdS black holes, respectively, the relations of  the shadow radius $r_{\rm sh}$, the deformation parameters  $\delta _{s}$ and  $k_{s}$, and  the circularity deviation  $\Delta C$  with respect to the horizon radius  $r_{\rm H}$  for  different inclination angles  $\theta _{0}$ at constant pressure $P$ and angular momentum $J$. We can see that $r_{\rm sh}$  monotonically increases, but  $\delta _{s}$, $k_{s}$  and  $\Delta C$  monotonically decrease with increasing of $r_{\rm H}$,  which  means
\begin{equation}
\label{96}
\frac{\mathrm{d} r_{\rm sh}}{\mathrm{d} r_{\rm H}}>0,  \qquad   \frac{\mathrm{d} (1/\delta _{s})}{\mathrm{d} r_{\rm H}}>0,   \qquad
\frac{\mathrm{d} (1/k _{s})}{\mathrm{d} r_{\rm H}}>0,  \qquad    \frac{\mathrm{d} (1/\Delta C)}{\mathrm{d} r_{\rm H}}>0.
\end{equation}

\begin{figure}[H]
		\centering
		\begin{minipage}{.5\textwidth}
			\centering
			\includegraphics[width=85mm]{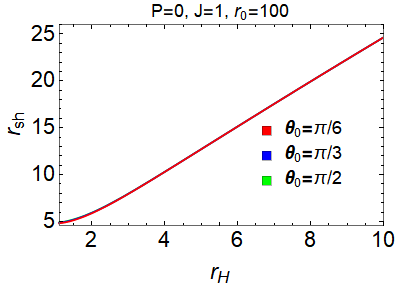}
		\end{minipage}%
		\begin{minipage}{.5\textwidth}
			\centering
			\includegraphics[width=85mm]{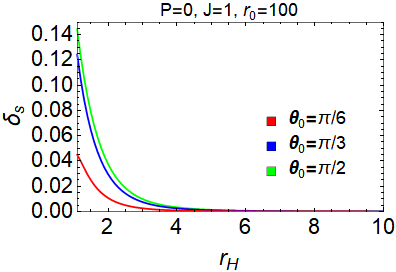}
		\end{minipage}
		\begin{minipage}{.5\textwidth}
			\centering
			\includegraphics[width=85mm]{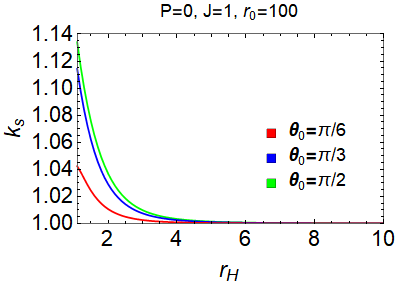}
		\end{minipage}%
		\begin{minipage}{.5\textwidth}
			\centering
			\includegraphics[width=85mm]{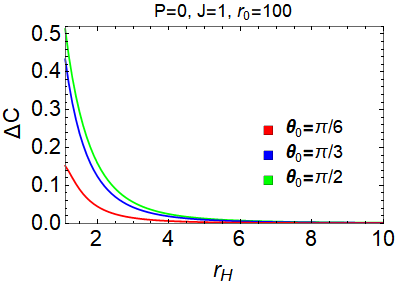}
		\end{minipage}		
		\caption*{Fig. 3. The shadow radius $r_{\rm sh}$, the deformation parameters  $\delta _{s}$ and  $k_{s}$, and  the circularity deviation  $\Delta C$  with respect to the horizon radius  $r_{\rm H}$  at constant pressure $P$ and angular momentum $J$  for the Kerr  black hole.}
\label{figure3}
\end{figure}

\begin{figure}[H]
		\centering
		\begin{minipage}{.5\textwidth}
			\centering
			\includegraphics[width=85mm]{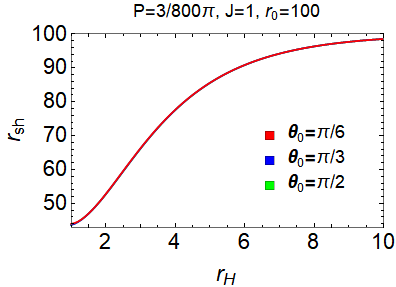}
		\end{minipage}%
		\begin{minipage}{.5\textwidth}
			\centering
			\includegraphics[width=85mm]{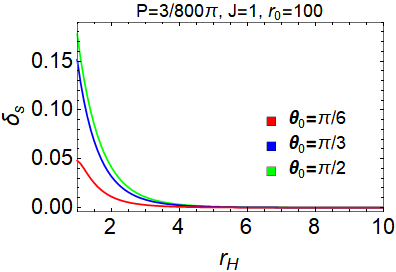}
		\end{minipage}
		\begin{minipage}{.5\textwidth}
			\centering
			\includegraphics[width=85mm]{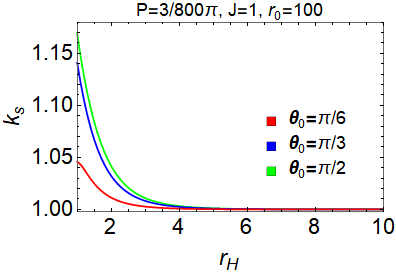}
		\end{minipage}%
		\begin{minipage}{.5\textwidth}
			\centering
			\includegraphics[width=85mm]{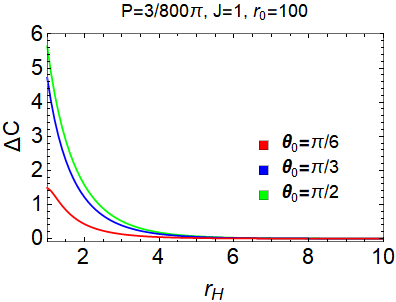}
		\end{minipage}		
		\caption*{Fig. 4. The shadow radius $r_{\rm sh}$, the deformation parameters  $\delta _{s}$ and  $k_{s}$, and  the circularity deviation  $\Delta C$  with respect to the horizon radius  $r_{\rm H}$  at constant pressure $P$ and angular momentum $J$  for the Kerr AdS black hole.}
\label{figure4}
\end{figure}

Considering $\frac{\partial T}{\partial r_{\rm H}}=\frac{\partial T}{\partial r_{\rm sh}}\frac{\mathrm{d} r_{\rm sh}}{\mathrm{d}r_{\rm H}}$,
$\frac{\partial R_{\rm SJ}}{\partial r_{\rm H}}=\frac{\partial  R_{\rm SJ}}{\partial r_{\rm sh}}\frac{\mathrm{d} r_{\rm sh}}{\mathrm{d}r_{\rm H}}$, and  Eq.~(\ref{96}),
we find that  the positivity or negativity of $\frac{\partial T}{\partial r_{\rm H}}$  and  $\frac{\partial R_{\rm SJ}}{\partial r_{\rm H}}$  depends entirely on the positivity or negativity of  $\frac{\partial T}{\partial r_{\rm sh}}$  and  $\frac{\partial  R_{\rm SJ}}{\partial r_{\rm sh}}$, respectively.  As a result, the conditions
\begin{equation}
\label{97}
\frac{\partial T}{\partial r_{\rm H}}>0, \qquad   \frac{\partial T}{\partial r_{\rm H}}=0, \qquad   \frac{\partial T}{\partial r_{\rm H}}<0,
\end{equation}
and
\begin{equation}
\label{98}
\frac{\partial R_{\rm SJ}}{\partial r_{\rm H}}>0, \qquad   \frac{\partial R_{\rm SJ}}{\partial r_{\rm H}}=0, \qquad   \frac{\partial R_{\rm SJ}}{\partial r_{\rm H}}<0,
\end{equation}
can be converted to
\begin{equation}
\label{99}
\frac{\partial T}{\partial r_{\rm sh}}>0, \qquad   \frac{\partial T}{\partial r_{\rm sh}}=0, \qquad   \frac{\partial T}{\partial r_{\rm sh}}<0,
\end{equation}
\begin{equation}
\label{100}
\frac{\partial T}{\partial (1/\delta _{s})}>0, \qquad   \frac{\partial T}{\partial (1/\delta _{s})}=0, \qquad   \frac{\partial T}{\partial (1/\delta _{s})}<0,
\end{equation}
\begin{equation}
\label{101}
\frac{\partial T}{\partial (1/k _{s})}>0, \qquad   \frac{\partial T}{\partial  (1/k _{s})}=0, \qquad   \frac{\partial T}{\partial (1/k _{s})}<0,
\end{equation}
\begin{equation}
\label{102}
\frac{\partial T}{\partial (1/\Delta C)}>0, \qquad   \frac{\partial T}{\partial (1/\Delta C)}=0, \qquad   \frac{\partial T}{\partial (1/\Delta C)}<0,
\end{equation}
and
\begin{equation}
\label{103}
\frac{\partial R_{\rm SJ}}{\partial r_{\rm sh}}>0, \qquad   \frac{\partial R_{\rm SJ}}{\partial r_{\rm sh}}=0, \qquad   \frac{\partial R_{\rm SJ}}{\partial r_{\rm sh}}<0,
\end{equation}
\begin{equation}
\label{104}
\frac{\partial R_{\rm SJ}}{\partial (1/\delta _{s})}>0, \qquad   \frac{\partial R_{\rm SJ}}{\partial (1/\delta _{s})}=0, \qquad   \frac{\partial R_{\rm SJ}}{\partial (1/\delta _{s})}<0,
\end{equation}
\begin{equation}
\label{105}
\frac{\partial R_{\rm SJ}}{\partial (1/k _{s})}>0, \qquad   \frac{\partial R_{\rm SJ}}{\partial  (1/k _{s})}=0, \qquad   \frac{\partial R_{\rm SJ}}{\partial (1/k _{s})}<0,
\end{equation}
\begin{equation}
\label{106}
\frac{\partial R_{\rm SJ}}{\partial (1/\Delta C)}>0, \qquad   \frac{\partial R_{\rm SJ}}{\partial (1/\Delta C)}=0, \qquad   \frac{\partial R_{\rm SJ}}{\partial (1/\Delta C)}<0.
\end{equation}

In order to test whether the relations Eqs.~(\ref{97})-(\ref{106}) are valid or not, according to Eqs.~(\ref{92})-(\ref{95}) we  draw  the Hawking temperature  $T$ and  the Ruppeiner thermodynamic scalar curvature $R_{\rm SJ}$ of  the Kerr (AdS)  black hole
as a function of $r_{\rm sh}$,  or $1/\delta _{s}$, or $1/k_{s}$,  or $1/\Delta C$, respectively,  at constant pressure $P$ and  angular momentum $J$  in Figs. 5-8,  where  the red curves  $T-r_{\rm H}$ (Eq.~(\ref{565})) and   $R_{\rm SJ}-r_{\rm H}$ (Eq.~(\ref{57}))   are  attached  as a comparison.

\begin{figure}[H]
		\centering
		\begin{minipage}{.5\textwidth}
			\centering
			\includegraphics[width=85mm]{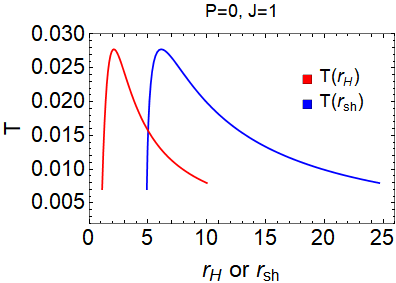}
		\end{minipage}%
		\begin{minipage}{.5\textwidth}
			\centering
			\includegraphics[width=85mm]{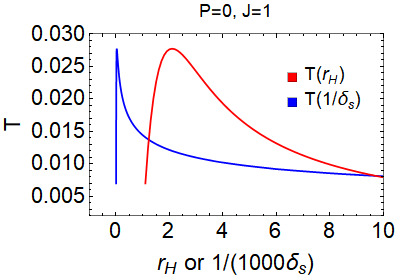}
		\end{minipage}
		\begin{minipage}{.5\textwidth}
			\centering
			\includegraphics[width=85mm]{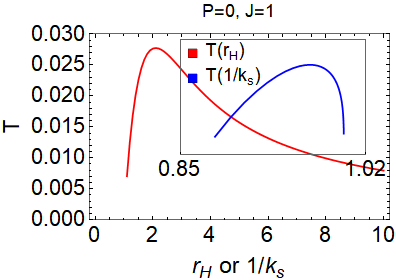}
		\end{minipage}%
		\begin{minipage}{.5\textwidth}
			\centering
			\includegraphics[width=85mm]{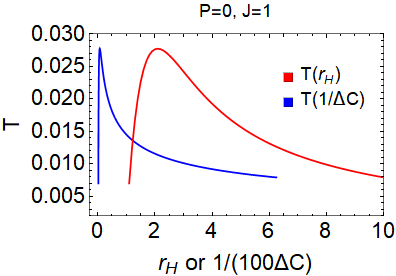}
		\end{minipage}		
		\caption*{Fig. 5. The Hawking temperature  $T$  of  the Kerr  black hole  as a function of $r_{\rm sh}$,  or $1/\delta _{s}$, or $1/k_{s}$,  or  $1/\Delta C$  at constant pressure $P$ and angular momentum $J$, where the red curves of the Hawking temperature  $T$  of  the Kerr  black hole  as a function of $r_{\rm H}$ are attached for comparison. Here  we  set  $r_{\rm 0}=100$  and  $\theta _{\rm 0}=\frac{\pi }{2}$.}
\label{figure5}
\end{figure}

\begin{figure}[H]
		\centering
		\begin{minipage}{.5\textwidth}
			\centering
			\includegraphics[width=85mm]{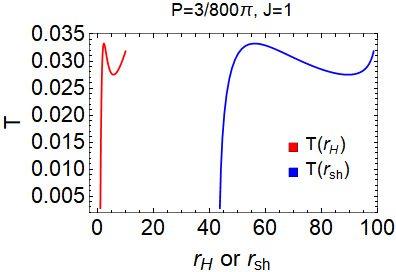}
		\end{minipage}%
		\begin{minipage}{.5\textwidth}
			\centering
			\includegraphics[width=85mm]{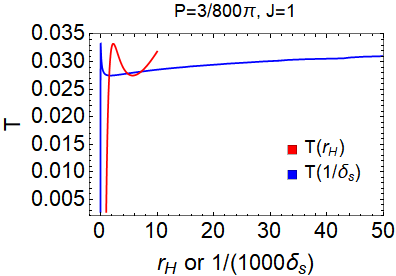}
		\end{minipage}
		\begin{minipage}{.5\textwidth}
			\centering
			\includegraphics[width=85mm]{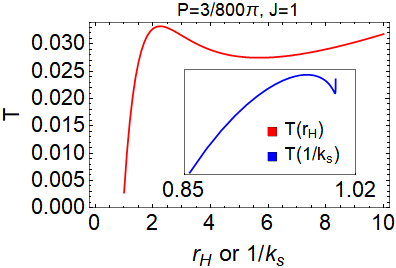}
		\end{minipage}%
		\begin{minipage}{.5\textwidth}
			\centering
			\includegraphics[width=85mm]{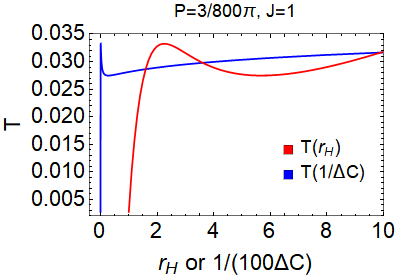}
		\end{minipage}		
		\caption*{Fig. 6. The Hawking temperature  $T$  of  the Kerr AdS black hole  as a function of $r_{\rm sh}$,  or $1/\delta _{s}$, or $1/k_{s}$,  or  $1/\Delta C$  at constant pressure $P$ and angular momentum $J$, where the  red curves of the Hawking temperature  $T$  of  the Kerr  AdS   black hole  as a function of $r_{\rm H}$ are attached for comparison. Here  we  set  $r_{\rm 0}=100$  and  $\theta _{\rm 0}=\frac{\pi }{2}$.}
\label{figure6}
\end{figure}

\begin{figure}[H]
		\centering
		\begin{minipage}{.5\textwidth}
			\centering
			\includegraphics[width=85mm]{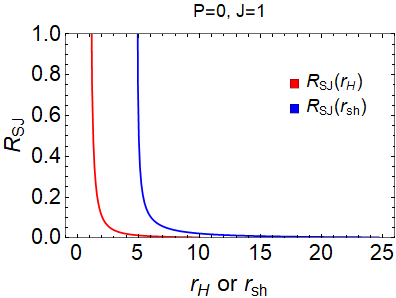}
		\end{minipage}%
		\begin{minipage}{.5\textwidth}
			\centering
			\includegraphics[width=85mm]{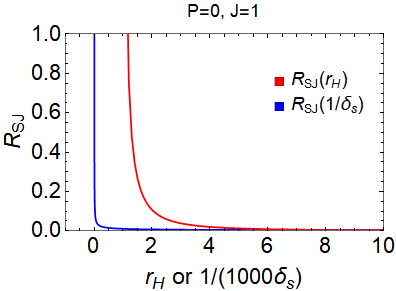}
		\end{minipage}
		\begin{minipage}{.5\textwidth}
			\centering
			\includegraphics[width=85mm]{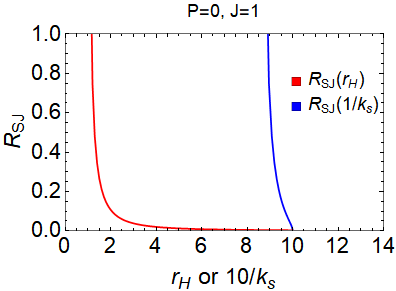}
		\end{minipage}%
		\begin{minipage}{.5\textwidth}
			\centering
			\includegraphics[width=85mm]{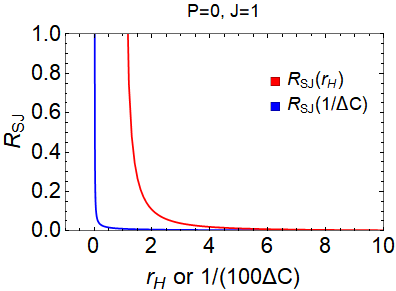}
		\end{minipage}		
		\caption*{Fig. 7. The Ruppeiner thermodynamic scalar curvature $R_{\rm SJ}$  of  the Kerr  black hole  as a function of $r_{\rm sh}$,  or $1/\delta _{s}$, or $1/k_{s}$,  or  $1/\Delta C$  at constant pressure $P$ and angular momentum $J$, where the red curves of the Ruppeiner  curvature $R_{\rm SJ}$  of  the Kerr  black hole  as a function of $r_{\rm H}$ are attached for comparison.  Here  we  set  $r_{\rm 0}=100$  and  $\theta _{\rm 0}=\frac{\pi }{2}$.}
\label{figure7}
\end{figure}

\begin{figure}[H]
		\centering
		\begin{minipage}{.5\textwidth}
			\centering
			\includegraphics[width=85mm]{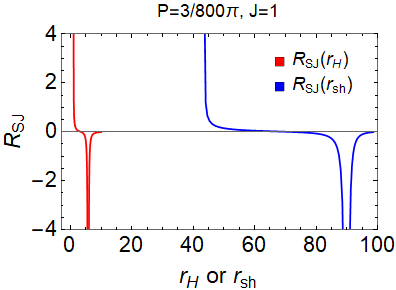}
		\end{minipage}%
		\begin{minipage}{.5\textwidth}
			\centering
			\includegraphics[width=85mm]{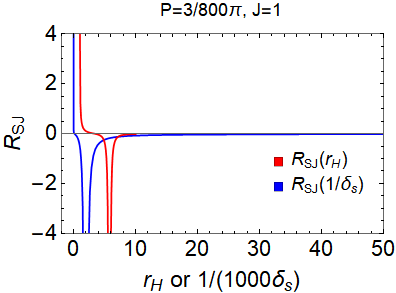}
		\end{minipage}
		\begin{minipage}{.5\textwidth}
			\centering
			\includegraphics[width=85mm]{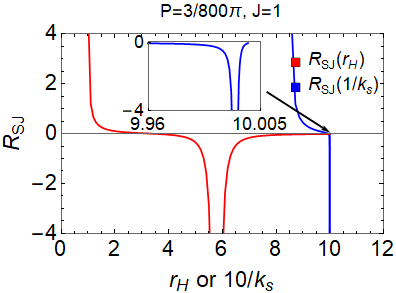}
		\end{minipage}%
		\begin{minipage}{.5\textwidth}
			\centering
			\includegraphics[width=85mm]{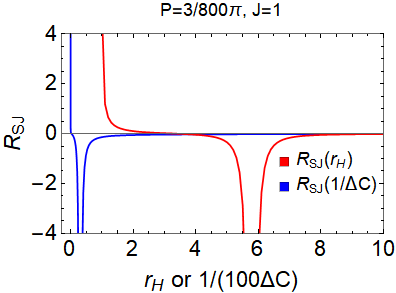}
		\end{minipage}		
		\caption*{Fig. 8. The Ruppeiner thermodynamic scalar curvature $R_{\rm SJ}$  of  the Kerr AdS  black hole  as a function of $r_{\rm sh}$,  or $1/\delta _{s}$, or $1/k_{s}$,  or $1/\Delta C$  at constant pressure $P$ and angular momentum $J$, where the red  curves of the Ruppeiner  curvature $R_{\rm SJ}$  of  the Kerr  AdS  black hole  as a function of $r_{\rm H}$ are attached for comparison. Here  we  set  $r_{\rm 0}=100$  and  $\theta _{\rm 0}=\frac{\pi }{2}$.}
\label{figure8}
\end{figure}

By comparing the blue curves and red curves  in  Figs. 5 and 6, we can clearly see that the blue curves  $T-r_{\rm sh}$, $T-1/\delta _{s}$,  $T-1/k _{s}$,   and  $T-1/\Delta C$  have similar profiles with the red  curve  $T-r_{\rm H}$,  which indicates that  the shadow radius $r_{\rm sh}$, the deformation parameters  $\delta _{s}$ and $k_{s}$, and the circularity deviation  $\Delta C$  can connect to the phase  transition of  black  holes.  Moreover, we can see clearly in Figs. 7 and 8 that  the blue  curves $R_{\rm SJ}-r_{\rm sh}$, $R_{\rm SJ}-1/\delta _{s}$, $R_{\rm SJ}-1/k _{s}$,  and  $R_{\rm SJ}-1/\Delta C$    have similar profiles with the red  curve   $R_{\rm SJ}-r_{\rm H}$,  which indicates that  the shadow radius $r_{\rm sh}$, the deformation parameters  $\delta _{s}$ and $k_{s}$, and the circularity deviation  $\Delta C$  can connect to the   black  hole  microstructure.  Here we note that our choice of  shadow deformation parameters comes from Refs.~\cite{PDD32,PDD33} but not from Ref.~\cite{P27} because the former is able to reflect the thermodynamics of the Kerr (AdS) black hole, while the latter is not.

Finally, we point out that the equivalence between  Eqs.~(\ref{565})-(\ref{57}) and Eqs.~(\ref{97})-(\ref{106}) is also valid  for the Kerr dS black hole with negative pressure. We verify this result by  depicting the relations of  $r_{\rm sh}$,   $\delta _{s}$, $k_{s}$, and  $\Delta C$ with respect to  $r_{\rm H}$  at different inclination angles  $\theta _{0}$  but at fixed pressure $P$ and angular momentum $J$  in Fig. 9.
We can see that $r_{\rm sh}$  monotonically increases but $\delta _{s}$,  $k_{s}$,  and  $\Delta C$  monotonically decrease  with increasing of $r_{\rm H}$,  which  is similar to the situation in Fig. 4.  As a result,  we conclude that the four observables, $r_{\rm sh}$,   $\delta _{s}$, $k_{s}$, and  $\Delta C$,  describing  the characteristics of shadows are  good enough to  have a close connection to the thermodynamics of Kerr (A)dS black holes.

\begin{figure}[H]
		\centering
		\begin{minipage}{.5\textwidth}
			\centering
			\includegraphics[width=85mm]{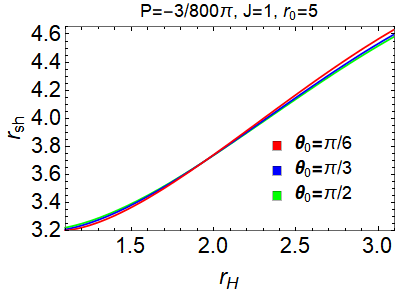}
		\end{minipage}%
		\begin{minipage}{.5\textwidth}
			\centering
			\includegraphics[width=85mm]{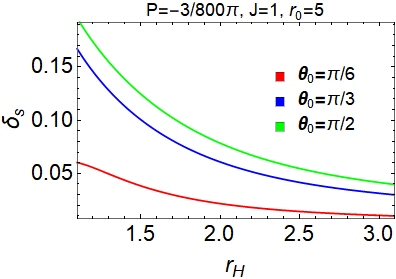}
		\end{minipage}
		\begin{minipage}{.5\textwidth}
			\centering
			\includegraphics[width=85mm]{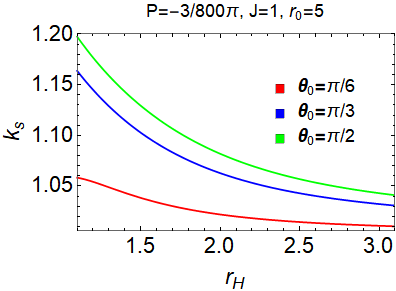}
		\end{minipage}%
		\begin{minipage}{.5\textwidth}
			\centering
			\includegraphics[width=85mm]{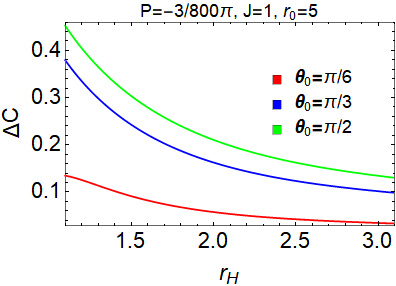}
		\end{minipage}		
		\caption*{Fig. 9. The shadow radius $r_{\rm sh}$, the deformation parameters  $\delta _{s}$ and $k_{s}$, and  the circularity deviation  $\Delta C$ with respect to the horizon radius  $r_{\rm H}$  at constant pressure $P$ and angular momentum $J$  for the Kerr dS black hole.}
\label{figure9}
\end{figure}

\section{Constraint to  the relaxation time of  black hole perturbation  based on  shadow data}

In this section, we use the shadow data to impose a constraint on the relaxation time defined as the period that a perturbed black hole returns to its stable configuration.
Based on   the  principle of  causality   and   the  second law of thermodynamics,  Bekenstein proposed~\cite{PDD35,PDD36} the  universal  bound  on the information emission rate of a physical system.
Later,  in terms of the laws of black hole thermodynamics,  Hod  developed Bekenstein's idea and derived~\cite{PDD37,PMM1} a  universal  bound  on  the relaxation time of  a perturbed  thermodynamic system, which is known as the Bekenstein-Hod universal bound,
\begin{equation}
\label{108}
\tau\geq \frac{\hbar}{\pi k_{\rm B}T},
\end{equation}
where $\tau $  represents  the relaxation time and can be regarded~\cite{PDD37,PMM1} as $1/|\omega _{\rm I}|$, and $|\omega _{\rm I}|$ denotes the absolute value of  imaginary part of fundamental and least damped quasinormal resonances. Moreover,
the  Bekenstein-Hod universal bound  is also related to~\cite{HHDD1,HHDD2,HHDD3} the viscosity-entropy bound.  Based on the remnant of binary black hole merger  observed by LIGO and  Virgo~\cite{PDD40,PDD41}, this bound  has recently been tested~\cite{PDD38} by the astrophysical black hole population  with $94\%$ probability.

According to the latest shadow data~\cite{P4,PDD23,PDD24,PDD25} of  the M$87^{*}$ black hole,  the angular size of the shadow  is $\delta =(42\pm 3)$ $\mu as$ (microarcsecond), the distance to the  black  hole  $D=16.8_{-0.8}^{+0.8}$ Mpc (million parsec), and the black hole mass $M=(6.5\pm 0.7)\times 10^{9}M_{\bigodot}$. As a consequence, the shadow diameter $d_{{\rm M}87^{*}}$  in terms of the unit of mass  is~\cite{PDD23,PDD24}
\begin{equation}
\label{1003}
d_{{\rm M}87^{*}}\equiv \frac{D\delta }{M}\approx 11.0\pm 1.5,
\end{equation}
which means that  the range of  the shadow diameter is  $9.5 \lesssim d_{{\rm M}87^{*}}=2r_{\rm sh}\lesssim  12.5$  in the  $1\sigma $ confidence region. In addition, the range of  the circularity deviation  $\Delta C$ of the  M$87^{*}$ shadow reads~\cite{P4}:  $\Delta C\leqslant 0.1$.
If the M$87^{*}$  black  hole is regarded as the Reissner-Nordstr\"om (RN) black hole, i.e., matching the shadow data to the RN  black  hole, we have the  available ranges  of charge $Q$ and Hawking temperature  $T$;  if  the M$87^{*}$  black  hole is regarded as the Kerr black hole, i.e., matching the shadow data to suit the Kerr black  hole, we have  the available  ranges of  angular momentum $J$ and Hawking temperature  $T$. These ranges are shown  in the left and right diagrams of Fig. 10, respectively.


\begin{figure}
		\centering
		\begin{minipage}{.5\textwidth}
			\centering
			\includegraphics[width=85mm]{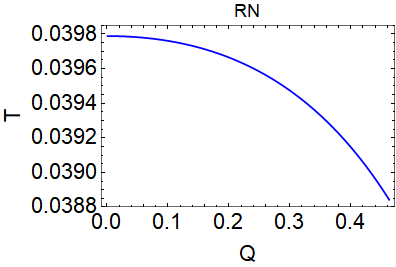}
		\end{minipage}%
		\begin{minipage}{.5\textwidth}
			\centering
			\includegraphics[width=85mm]{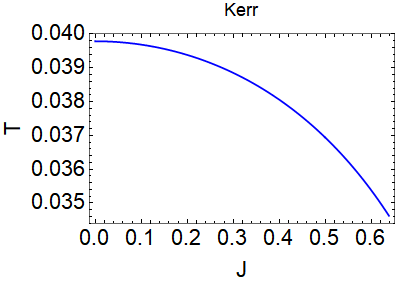}
		\end{minipage}
\caption*{Fig. 10.  The Hawking temperature  $T$ with respect to the charge $Q$ in the left diagram if matching the shadow data of  the  M$87^{*}$  black  hole to suit  the RN black  hole, and the Hawking temperature  $T$ with respect to the angular momentum $J$ in the right diagram if matching the shadow data of  the  M$87^{*}$  black  hole to suit the Kerr black  hole. Here we set  $M=1$.}
\label{figure10}
\end{figure}

Combining Fig. 10 with  Eq.~(\ref{108}), we can deduce the value range of  the  relaxation time  $\tau$  for the RN black hole and  the Kerr black hole, respectively, when matching the shadow data of the
M$87^{*}$  black  hole  to suit them,
\begin{equation}
\label{110}
\tau^{\rm RN}\geq \tau _{\rm min}^{\rm RN},  \qquad  \tau _{\rm min}^{\rm RN}\in [8, 8.2],
\end{equation}
\begin{equation}
\label{111}
\tau^{\rm Kerr}\geq \tau _{\rm min}^{\rm Kerr},   \qquad    \tau _{\rm min}^{\rm Kerr}\in [8, 9.2].
\end{equation}

From Fig. 10 we can see that the increase of charge $Q$ ($Q\equiv q^2$)  and  angular momentum $J$ will make the Hawking temperature $T$ decrease when the black hole  mass $M$  is fixed,  which  is shown by the blue curves.  This implies that the Schwarzschild black hole has the largest Hawking temperature  $T_{\rm max}=\frac{\hbar c^{3}}{8\pi k_{\rm B}GM}$  at  a  fixed  black hole mass.
Therefore, by substituting the largest  Hawking temperature  $T_{\rm max}$ into  Eq.~(\ref{108}), one can obtain~\cite{HHDD3} the minimum relaxation time  at  a  fixed  black hole mass  $M$,
\begin{equation}
\label{20211}
\tau _{\rm min}=\frac{8GM}{c^{3}}.
\end{equation}
Using Eq.~(\ref{20211}), we  know that  the relaxation time for  the perturbed M$87^{*}$ black hole to return to its equilibrium  equals at least,
\begin{equation}
\label{112}
 \tau _{\rm min}^{{\rm M}87^{*}}=\frac{8GM_{{\rm M}87^{*}}}{c^{3}}=256149\; {\rm seconds}\approx 3\;{\rm days},
\end{equation}
where  $M_{{\rm M}87^{*}}=6.5\times 10^{9}M_{\bigodot }$. Similarly,  we obtain the corresponding minimum relaxation time  of the  Sgr $A^{*}$  black  hole,
\begin{equation}
\label{113}
 \tau _{\rm min}^{{\rm Sgr}\, A^{*}}=\frac{8GM_{{\rm Sgr}\, A^{*}}}{c^{3}}=158.42\;{\rm seconds}\approx 2.64\;{\rm minutes},
\end{equation}
where the mass of the supermassive black hole in the center of the Milky Way is~\cite{PDD39}: $M_{{\rm Sgr}\, A^{*}}=4.02\times 10^{6}M_{\bigodot }$.

In addition, we can know that the Schwarzschild black hole has~\cite{PDD7,PHLF1} the largest shadow size at a fixed black hole mass $M$,
\begin{equation}
\label{118}
 r_{\rm sh}^{\rm max}=\frac{3\sqrt{3}GM}{c^{2}}.
\end{equation}
Using Eqs.~(\ref{20211})  and  (\ref{118}),  we obtain the linear relationship between  the minimum relaxation time  $\tau _{\rm min}$  and the maximum shadow radius  $ r_{\rm sh}^{\rm max}$,
\begin{equation}
\label{713}
\tau _{\rm min}=\frac{8\sqrt{3}}{9c}r_{\rm sh}^{\rm max},
\end{equation}
and  for the first time  plot the graph of the minimum relaxation time  $\tau _{\rm min}$  with respect to the maximum shadow radius  $ r_{\rm sh}^{\rm max}$ for different black hole mass levels in Fig. 11.

\begin{figure}[H]
		\centering
		\begin{minipage}{.8\textwidth}
			\centering
			\includegraphics[width=130mm]{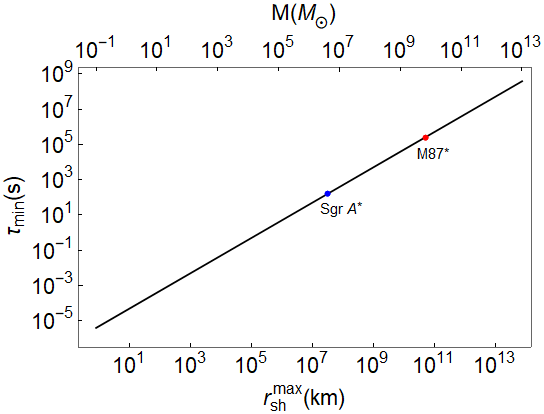}
		\end{minipage}%
\caption*{Fig. 11.  The minimum relaxation time  $\tau _{\rm min}$  with respect to  the maximum shadow radius  $ r_{\rm sh}^{\rm max}$  for different black hole mass levels. The coordinates of the red dot are  (6.5$\times$$ 10^{9}$, 4.98775$\times$$ 10^{10}$, 256149), which represents the mass, the maximum shadow radius, and the minimum relaxation time of the  M$87^{*}$  black  hole, while the coordinates of the blue dot are  (4.02$\times$$ 10^{6}$, 3.08473$\times$$ 10^{7}$, 158.42),  which represents the mass,  the  maximum shadow radius, and the minimum relaxation time  of the Sgr $A^{*}$  black hole.}
\label{figure11}
\end{figure}

We note that Eq.~(\ref{713}) has a certain universality for asymptotically flat black holes, but not for asymptotically non-flat black holes.
From Fig. 11, we draw the following conclusions:
\begin{itemize}
\item  When the black hole mass $M$ is fixed, there exists a linear relationship between the minimum relaxation time  $\tau _{\rm min}$ and  the maximum shadow radius
$ r_{\rm sh}^{\rm max}$,  where the former represents  the  thermodynamics of black holes   while the latter the dynamics of black holes. Quite interesting is that the thermodynamic effect is proportional to the dynamic one.
\item  Given the mass $M$ of a celestial black hole, we can know the minimum relaxation time  $\tau _{\rm min}$ and the maximum shadow radius   $ r_{\rm sh}^{\rm max}$ in terms of this figure. For example, using the mass of the  M$87^{*}$  black  hole and the mass of the Sgr $A^{*}$  black hole, we can mark the minimum relaxation time and  the maximum shadow radius of the two black holes in this figure. The two points are located on a line. Using this line, we can predict the minimum relaxation time  and  the maximum shadow radius for a new black hole when its mass is measured.
\end{itemize}
We look forward to verifying Fig. 11 by using more shadow data and   gravitational wave data of celestial black holes from astronomical observations in the future.

\section{Conclusion}

In this paper, at first we investigate the relationship between the shadow radius and  microstructure for a general static spherically symmetric black hole, and then turn to the relationship between the shadow  and  thermodynamics  in the aspects of phase transition and microstructure for the Kerr (AdS) black hole.  Moreover, we  give the constraint to  the  relaxation  time  of  black hole perturbation by combining the shadow data  with  the Bekenstein-Hod universal bound.  We summarize our main results as follows.
\begin{itemize}
\item   The shadow radius of a general static spherically symmetric black hole can reflect the  black hole  microstructure.
\item  The four observables, the shadow radius $r_{\rm sh}$, the deformation parameters  $\delta _{s}$ and  $k_{s}$, and  the circularity deviation  $\Delta C$, describing the shadow characteristics of the Kerr (A)dS black hole can reflect the black hole thermodynamics  in the aspects of phase transition and microstructure.
\item  We  predict that  the minimum relaxation times  of   M$87^{*}$  black  hole  and  Sgr $A^{*}$ black hole  are approximately 3 days and  2.64 minutes, respectively.  Moreover,  we plot Fig. 11 describing the relation among the three parameters, $(M, r_{\rm sh}^{\rm max}, \tau _{\rm min})$, on a plane,   and expect it to be verified  in the future.
\end{itemize}

In  a word,  there exists a close relationship between dynamics and thermodynamics of black holes, where the black hole shadow represents dynamics, the Hawking temperature $T$ describes phase transitions, and the thermodynamic curvature scalar $R$ reflects black hole microstructure.  This shows that the unobservable black hole thermodynamics  can indeed be revealed by the observable dynamics. Especially, the shadow is a good enough observable to unfold the black hole thermodynamics.

\section*{Acknowledgments}

This work was supported in part by the National Natural Science Foundation of China under Grant Nos. 11675081 and 12175108.


\end{document}